# Scholars on Twitter: who and how many are they?

Rodrigo Costas[1, 2], Jeroen van Honk[1], Thomas Franssen[1]

*rcostas@cwts.leidenuniv.nl*; *j.s.van.honk@cwts.leidenuniv.nl*; *t.p.franssen@cwts.leidenuniv.nl*
[1] Centre for Science and Technology Studies (CWTS) Leiden University, Leiden (the Netherlands)

[2] Centre for Research on Evaluation, Science and Technology (CREST), Stellenbosch University, Stellenbosch (South Africa)

**Abstract**
In this paper we present a novel methodology for identifying scholars with a Twitter account. By combining bibliometric data from Web of Science and Twitter users identified by Altmetric.com we have obtained the largest set of individual scholars matched with Twitter users made so far. Our methodology consists of a combination of matching algorithms, considering different linguistic elements of both author names and Twitter names; followed by a rule-based scoring system that weights the common occurrence of several elements related with the names, individual elements and activities of both Twitter users and scholars matched. Our results indicate that about 2% of the overall population of scholars in the Web of Science is active on Twitter. By domain we find a strong presence of researchers from the Social Sciences and the Humanities. Natural Sciences is the domain with the lowest level of scholars on Twitter. Researchers on Twitter also tend to be younger than those that are not on Twitter. As this is a bibliometric-based approach, it is important to highlight the reliance of the method on the number of publications produced and tweeted by the scholars, thus the share of scholars on Twitter ranges between 1% and 5% depending on their level of productivity. Further research is suggested in order to improve and expand the methodology.

**Conference Topic**
Altmetrics; Studies on the level of individual scientists

## Introduction

Social media have become increasingly important as a tool for scholarly communication (Sugimoto et al., 2016). Academic social networking sites such as Mendeley, Academia.edu and ResearchGate are used by scientists to connect with peers, share (pre)prints and track the visibility of their work. Twitter is a popular microblogging platform in which diverse communities, such as journalists, students and scientists among others, interact and exchange information. Earlier research has studied discourses on Twitter in comparison to news discourses (Mondragon et al., 2017) and the use of Twitter amongst different scientific disciplines (Holmberg & Thelwall, 2014). A major drawback of these previous studies is their small scale. The majority of research on Twitter use amongst academics deals with 500 accounts or less (but see Hadgu & Jäschke, 2014). In general, there is a lack of identification methods that can be used to identify scientists on Twitter on a larger scale.

Earlier work has used a variety of (labour intensive) methods to identify scientists on Twitter. The most commonly used method is based on selecting Twitter accounts manually (Veletsianos, 2012; Lulic & Kovic, 2012; Holmberg & Thelwall, 2014; Haustein et al., 2014; Hwong et al., 2016), often starting from an established set of scientists through snowball sampling (identifying for instance all scientists that follow or are followed by the established set of scientists). A second method is self-identification through survey research (Rowlands et al., 2011; Van Noorden, 2014; Collins, Shiffman, Rock, 2016). In this method, that potentially can be used to collect a larger group of scientists active on Twitter, researchers send out surveys to a group of scientists asking them, amongst others to identify their social media accounts. However, such method is always limited by response rate as well as the need to identify contact information for scientists before being able to contact them. A third method is by identifying scientists through (conference) hashtags or Twitter accounts (Ross et al., 2011; Hadgu & Jäschke, 2014; Veletsianos & Kimmons, 2016). Hadgu and Jäschke (2014) have been especially successful by combining accounts of various conferences to obtain a large sample of Twitter accounts affiliated with computer science conferences. A fourth method is the use of lists compiled by Twitter users (Sharma et al., 2012). This approach has resulted in the largest set of scientists on Twitter to date compiled by Ke, Ahn and Sugimoto (2016). This study collected Twitter accounts of lists where the list as well as the Twitter biography contained a scientist title (e.g. psychologist, economist, PhD, researcher, etc.). They were able to collect 45,867 Twitter accounts of (self-identified) scientists. A fifth strand of research on Twitter users originates in computer science analyses on the demography of Twitter. Such studies involve far less manual labor and often work with very large datasets extracted from the Twitter API. These studies use metadata of Twitter accounts (such as the Twitter biography and location) as well as tweets and the network of followers and following to identify geographical location, age, gender and occupation of users (Mislove et al., 2011; Sloan et al., 2013; Sloan et al., 2015). This approach is successful in obtaining information about the demographics of Twitter, but is not able to identify individual scientists. The inclusion in lists and the presence of scientists titles in the Twitter biography section, as developed by Ke, Ahn and Sugimoto (2016) is a viable strategy, however this approach favours more established scientists who are included in Twitter lists far more often than junior scientists.

In this paper we take a different approach by creating a match between two universes of individuals: the set of authors recorded in the Web of Science database and the Twitter accounts recorded by Altmetric.com. Altmetric.com has developed a database of all tweets (and their related metadata) that include a (link to) a DOI (mostly related to scientific journal articles), which gives us a population of 2,6 million Twitter accounts, many of them being very likely related to scholars. Our aim is to identify all those Twitter accounts that belong to individuals among the 22 million disambiguated authors recorded in the Web of Science with at least a publication after the year 2005.

We think this approach is more viable than the approaches described above for a number of reasons. First, drawing on these two datasets makes it possible to match scholars (determined by their authorship of publications covered in the Web of Science) and Twitter handles in a systematic manner, involving far less manual labour. Second, when we identify the Twitter account of an author, we can immediately connect this to other data (such as institution, field of activity, publication profile, citation profile, collaborations, etc.) related to the scientific author. Third, while it relies partly on self-identification (the author has to be on Twitter using –to some extent- her own real name and has to have tweeted a paper recorded in Altmetric.com) it uses a variety of factors to determine whether an account can be ascribed to a particular scholar, thus going beyond the mere identification of the Twitter user as a scholar in the Twitter biographical section. Fourth, we aim at validating the linkages between scholars and Twitter accounts based on a gold standard of valid set of scholars with Twitter accounts, for this using the self-reported Twitter accounts of researchers recorded in the public ORCID registry (https://orcid.org/).

**Methodology**

*Data sources*

The two data sources used for this matching are the Web of Science (WoS) database, and the Altmetric.com database. We use the author-name disambiguation algorithm developed by Caron & Van Eck (2014) and we work with a set of 22,642,206 disambiguated author names. From the Altmetric.com database we extracted all distinct Twitter accounts that have tweeted at least a paper recorded in Altmetric.com until April 2016 (i.e. 2,622,116 distinct Twitter handles).

*Matching author-Twitter names*

In order to match the two different data sets – WoS disambiguated authors and Twitter user accounts – we started with some normalisation and cleaning of the data in order to harmonize the characters from both sources. In WoS, Roman characters are always used. However, in Twitter, scholars can (and often do) also use other character sets. There is, as such, a bias toward this Roman character set inherent in the matching (although automatic transliteration of other character sets could be a future strand of research).

Moreover, accents and diacritics are also generally elided in the WoS database while not in Twitter. This has been compensated by applying specific matching strategies for specific accents. For instance, umlauts in German are often not merely removed, but rewritten with an added e, so that for instance *ü* becomes *ue*. In our matching we have accounted for both *ü, u* and *ue* in these cases. We have attempted to apply such flexible matching to all frequent diacritics and spelling variations. Another frequent example is allowing hyphens to match whitespace. These are often used interchangeably in double last names, and so we consider them interchangeable.

The data from the Twitter side has issues of its own too. We use two data fields recorded in the Altmetric.com database: the handle (or username), and the "full name" as the users have entered it in their Twitter account. Twitter makes no distinction between first and last name. In fact, sometimes users do not actually enter their real name in this field at all. However, we are not interested in these cases as we aim at researcher that quite clearly disclose their identity in their Twitter accounts. Still, when they provide their actual name, it can come in various guises. They can provide any combination of first name, initials, last name, maiden name, title and sometimes even institution, and these can occur in almost any order. By looking at the source data, we have tried to catch as many of these as possible within what we might call "format templates". So if we find a name that is formatted "X.X. Xxx" (where X is any alphabetical character) we assume that the first two characters are initials and the last three are the last name. If the name is formatted "XX Xxx" it will be the same, yet if the second X is in lowercase ("Xx Xxx") it will be considered a first name and last name. And so on. Another complication to this name field is that it is limited by Twitter to twenty characters. Of course, full names can easily exceed this, and it is hard to predict how users deal with this cut-off. In any case, it is sure to increase the number of incomplete names.

Matching of the Twitter account with the author name can also be done based on the name from the handle. Handles on Twitter have a maximum length of fifteen characters and can include only a combination of alphanumeric characters and underscores (_). Of course, the absence of a whitespace character as well as other characters that might appear in a name (like dots and dashes) makes this field less reliable than the full name field. On the other hand, the restrictions can also help. In the case of many Chinese and Japanese researchers, for instance, the full name field contains their name in their own character set, whereas in their Twitter username they have been forced to use the Romanized version. Like with the full name, we use "format templates" here. We here assumed underscores to function as dividers between first and last names, or initials and last name, and we also took into account the sequence of upper- and lowercase characters (so that the username "XxxYyy" can be divided into first name "Xxx" and last name "Yyy").

Based on these extraction processes and attempts to make matching as flexible as possible, the two data sets are finally connected and pairs of authors and Twitter users have been created.

*Scoring the matches*

Based on the previous matching approach, a total database with 503,599,561 pairs of author names and Twitter accounts has been created. The following step is to determine which of those pairs are correct. In order to do so, a scoring method based on different rules has been developed in order to select the most likely correct pairs of authors and Twitter accounts. This rule approach is inspired by a similar approach for author-name disambiguation developed by Caron & van Eck (2014). Several data elements that can happen both in Twitter data and in WoS data have been collected in order to weight the pairs. Different rules have been applied in order to score the occurrence of the different elements, and a final score by summing the different rule-based scores has been provided for each author-Twitter account pair. Table 1 presents a summary of the rules and their scores.

Table 1. Summary of the criteria and scores for the different elements matched.

| Rules | Matching event | Criteria | Score |
|---|---|---|---|
| 0 | General matching | This is the basic matching between authors and Twitter accounts. The minimum matching element required is surname and initial. Thus, by default, all matches have at least 1 score. | 1 |
| 1 | Full name of author and tweeter | Frequent full name (>30 scholars in the whole database with the same name) | 1 |
|  |  | Medium frequent full name (between 5 and 30 scholars in the whole database with the same name) | 2 |
|  |  | Infrequent full name (less than 5 scholars in the database with the same name) | 3 |
| 2 | First name | Frequent first name (>145 scholars in the whole database with the same first name) | 1 |
|  |  | Medium frequent first name (between 145 and 12 scholars in the whole database with the same first name) | 2 |
|  |  | Infrequent first name (less than 12 scholars in the database with the same first name) | 3 |
| 3 | First single name | Very frequent first single name (more than 31 scholars in the database with the same first single name) | 1 |
|  |  | Infrequent first single name (less than 31 scholars in the database with the same first single name) | 2 |
| *4* | *First single name penalization* | *When the author has a first name in the papers but such name does not appear in the Twitter name(s) at all.* | *-2[1]* |
| 5 | E-mail URL (in the Twitter account and as obtained from the e-mail server URL of the author) | Very frequent author URL (>187 scholars are linked to the same e-mail server URL) | 1 |
|  |  | Medium frequent author URL (between 187 and 18 scholars are linked to the same e-mail URL) | 2 |
|  |  | Infrequent author URL (less than 18 scholars are linked to the same e-mail URL) | 3 |
| 6 | Organization name | Very frequent organization name (>403 scholars are linked to the same organization) | 1 |
|  |  | Medium frequent organization name (between 403 and 20 scholars are linked to the same organization) | 2 |
|  |  | Infrequent organization name (less than 20 scholars are linked to the same organization) | 3 |
| 7 | City | Very frequent city (>5,515 scholars in the database are linked to an affiliation in the city) | 1 |
|  |  | Medium frequent city (between 5,515 and 210 scholars in the database are linked to an affiliation in the city) | 2 |
|  |  | Infrequent city (less than 210 scholars are linked in the database to | 3 |

---

[1] We penalize when an author uses her first name in her papers but uses a different one (or none) in the Twitter name.

| Rules | Matching event | Criteria | Score |
|---|---|---|---|
| | | an a affiliation in the city) | |
| 8 | Country | Very frequent country (country has more than 76,742 scholars in the database linked to them) | 1 |
| | | Medium frequent country (country has less than 76,742 scholars in the database linked to them) | 2 |
| 9 | Tweeter has tweeted publications from the author (i.e. self-tweeting) | n. pubs self-tweeted: 1-2 | 3 |
| | | n. pubs self-tweeted: 3-5 | 5 |
| | | n. pubs self-tweeted: >5 | 7 |
| 10 | Twitter user has tweeted publications from the same micro-topic(s)[2] of author's activity (excluding self-tweeting) | n. overlapping topics tweeted: 1-3 | 1 |
| | | n. topics tweeted: 4-6 | 3 |
| | | n. overlapping topics tweeted: >6 | 5 |
| 11 | Paired by co-tweeted | The tweeter has been mentioned in at least the same tweet with the paper of the author simultaneously | 5 |
| 12 | Tweeter has tweeted publications from the same journal(s) of author's activity (excluding self-tweeting) | n. journals 1-5 | 1 |
| | | n. journals >5 | 2 |
| 13 | Commonness of the Twitter account-researcher combination | Combination of 1-2 scholars/Twitter | 2 |
| | | Match with 3-6 scholars/Twitter | 1 |
| | | Match with >6 scholars/Twitter combination | 0 |
| 'Preferred rule' | When an author is paired with several Twitter accounts, by definition the author will be linked to the account with the highest score (as calculated with the rules above). In case of ties in the scores, all the matches are considered. | | No score |

As presented in table 1, there are a total of 14 rules, one of them with a negative score. The rules can be organized in four main groups.

1) <u>Rules based on the matching of the names of the authors and the Twitter names</u> (rules 0 to 4). These rules are based on a relative clear identification of both the authors and the Twitter users, and their matching. Rule [0] refers the general matching of authors and Twitter information done in the previous step. Thus, all potential matches get, in principle, a score of 1, although later on this can be changed with the negative rule.

2) <u>Rules based on the matching of individual-related information of the authors and the Twitter users</u> (rules 5 to 8). These rules are based on the matching of different elements (mostly institutional and geographical) provided both by the authors in their papers (e.g. affiliations, countries, e-mails) and the Twitter users (URLs, Twitter name and Twitter handle, geographical information provided in the Twitter profiles, biographical description in Twitter, sometimes including their city, country or also academic institution). In these rules, different scores are established depending on the frequency of the different elements in WoS. Thus, the higher the number of scholars that are linked to the same entity in WoS, the lower the score that will be attribute to that match.

3) <u>Rules based on the matching of activity-related information</u> (rules from 9 to 12). Here the rules are based on the publications, fields and journals of the authors and the publications tweeted by the Twitter account. The rationale is that the more a Twitter account (sharing a minimum name similarity with a given author) has tweeted the same papers of the author (i.e. 'self-tweeting' activity), papers from the same micro-fields, and/or from the same journals of activity of the author, the higher the chances that the author and the Twitter account correspond to the same person. Additionally, if a Twitter account has been mentioned (by another Twitter user) together with a paper of the author with which it has been matched, it is very likely that the Twitter account corresponds to the author, in a sort of co-tweeted event (e.g., somebody in Twitter has tweeted a link to a paper and the handle of the author in the same tweet[3]).

---
[2] Micro-topics are defined as the fields obtained in the publication-level classification developed by (Waltman & Van Eck, 2012)

These rules have the highest scores. Activity rules are expectedly more accurate in the identification and discrimination of the correct matches as they provide information on more specific patterns of the relationship between the scholar and the Twitter account.

There is also an additional rule [13] that assesses the commonness of the matching. If an author is only matched to one Twitter account, the matching is weighted more positively than when the author is matched to numerous different Twitter accounts.

Based on the previous rules, there is also an additional rule, labelled as the 'preferred rule'. It is not a scoring rule but a selection one. After performing the scoring of pairs of matched based on rules 0-13, each author (when matched to several Twitter accounts) is only assigned to the Twitter account with which it has the highest score. In case of ties (i.e. when an author is matched to several accounts with the same score) the author is kept linked to all of them. This rule is expected to reduce the noise caused by matches of authors with very similar names and activity profiles, working in the same institute, field, etc., which may cause that they get high scores in their combination with different Twitter accounts. It is a rule to increase the precision of the matching procedure.

*Gold standard, validation of the matching and final selection of author-Twitter handle pairs*

In order to determine the threshold of the score for the final selection of the most adequate matches, we have performed precision-recall analysis using a database of self-reported linkages of authors with their Twitter accounts. This 'gold standard' is based on the public ORCID data from years 2015 and 2016 (Paglione et al, 2015; Haak et al, 2016).

From these databases we found a total of 768 of different individuals that are also authors in the database of disambiguated scholars and have reported a Twitter account in their public ORCID profile. We further selected those individuals with a Twitter account in Altmetric.com (i.e. researchers that have tweeted at least one paper) finding a total of 631 different individuals. This means that about 82% of the scholars with a Twitter account in the gold standard are recorded (at least once) in the Altmetric.com database, thus suggesting the potential relevance of this method in order to be able to identify scholars[4]. Of the 631 we further discarded 4 scholars as they had their last publication before the year 2005. A total of 627 individuals were finally considered in the analysis. Based on the pairs of author names and Twitter names and in the scoring of the matches as described in table 1 we estimated the precision/recall values for those researchers in the gold standard. The results of such analysis are presented in Table 2. Table 2 presents the overall results of the number of scholars that have been paired with a Twitter account with a given score value (from ≥3 to ≥6), together with the precision/recall values based on the gold standard.

Table 2. Analysis of the scoring system of the matches

| Total score | Distinct scholars matched with a Twitter account | *Precision* | *Recall* |
|---|---|---|---|
| ≥6 | 134604 | 97.0% | 72.3% |

---
[3] See an example here: https://Twitter.com/wmijnhardt/status/781245999545212930
[4] Somehow it supports the idea that scholars that are on Twitter tend, at some point, to tweet or re-tweet a scientific publication, thus entering into the realm of altmetric sources (Haustein, Bowman & Costas, 2016).

| Total score | Distinct scholars matched with a Twitter account | *Precision* | *Recall* |
|---|---|---|---|
| ≥5 | 180484 | 96.3% | 76.5% |
| ≥4 | 385024 | 94.6% | 82.5% |
| ≥3 | 951169 | 91.9% | 88.0% |
| ≥2 | 3451743 | 78.8% | 91.5% |

As it can be seen in Table 2, precision increases with higher scores, while recall decreases. Basically, with scores higher than 5 we reach values of precision higher than 95%, but values of recall below 80%. Regarding recall, with a score low as 2 we get a value of 92%. In the discussion of recall it is important to keep in mind that not all the author-Twitter account recorded in ORCID are valid matches, for examples, there are Twitter accounts reported that are institutional or collective (e.g. labs, teams, departments, etc.). Therefore, the values of recall reported here have to be regarded as relatively conservative.

In this paper, in order to make a selection with a relative balance between precision and recall, and considering that precision is something to be preferred over recall (it is reasonable to argue that is best to have correct matches over more but noisy ones) a minimum value of 4 has been chosen as a good compromise. As a result, more than 385,000 individual scholars have been linked to a Twitter account. This number of scholars found in Twitter is the largest so far compared to any of the previous studies. Even if we would have selected a minimum score of 6, we would have obtained a set of almost 135,000 scholars on Twitter (i.e. almost three times more than in the largest study of identification of scholars on Twitter so far, cf. Ke, Ahn & Sugimoto, 2016).

**Analysis of the presence of scientists on Twitter**

In this section a general analysis on the presence of scholars with a Twitter account is presented. For this analysis we have focused on distinguishing those scholars that have a Twitter account as identified with the methodology explained above in contrast to those without a Twitter account. Only individuals with an article, review and letter and in the period 1980-2015 and with at least a publication after 2005 have been considered in the analysis (a total of 17,332,510 disambiguated scholars). A total of 334,856 (2%) of them have been matched with a Twitter account with a score of 4 or higher.

In Figure 1 the share of researchers that have a Twitter account controlling by production is presented. As it can be seen the share of scholars on Twitter increases with the number of publications, ranging between 1% and 5%. It could be argued that this is a sign that more research active scholars are also more likely to be active on social media. However, considering that some of the scoring rules (i.e. activity rules – 9,10 and 12) are dependent on the number of publications of the authors, the outcomes of the analysis could also indicate that researchers with more output are more likely to be linked to their Twitter account in our methodology.

In Figure 2 the share of scholars with a Twitter account by scientific domain is presented. In order to classify scholars into major scientific domains the same approach as used by Larivière & Costas (2016) has been employed here. Thus, individuals are assigned to their main domain of activity based on their number of publications, and in case of ties (i.e. a scholar linked to several domains with exactly the same number of publications) a random assignment is applied. Results show how scholars with Twitter accounts are prominent in disciplines such as 'Social and Behavioral Sciences' (above 5%) as well as 'Law, Arts and Humanities' (almost 4%). The shares in the 'Medical and Life Sciences' (about 2%) as well as in the 'Natural Sciences' (just above 1%) are much lower.

Figure 1. Overall number of scholars and percentage of them with a Twitter account, controlling by production

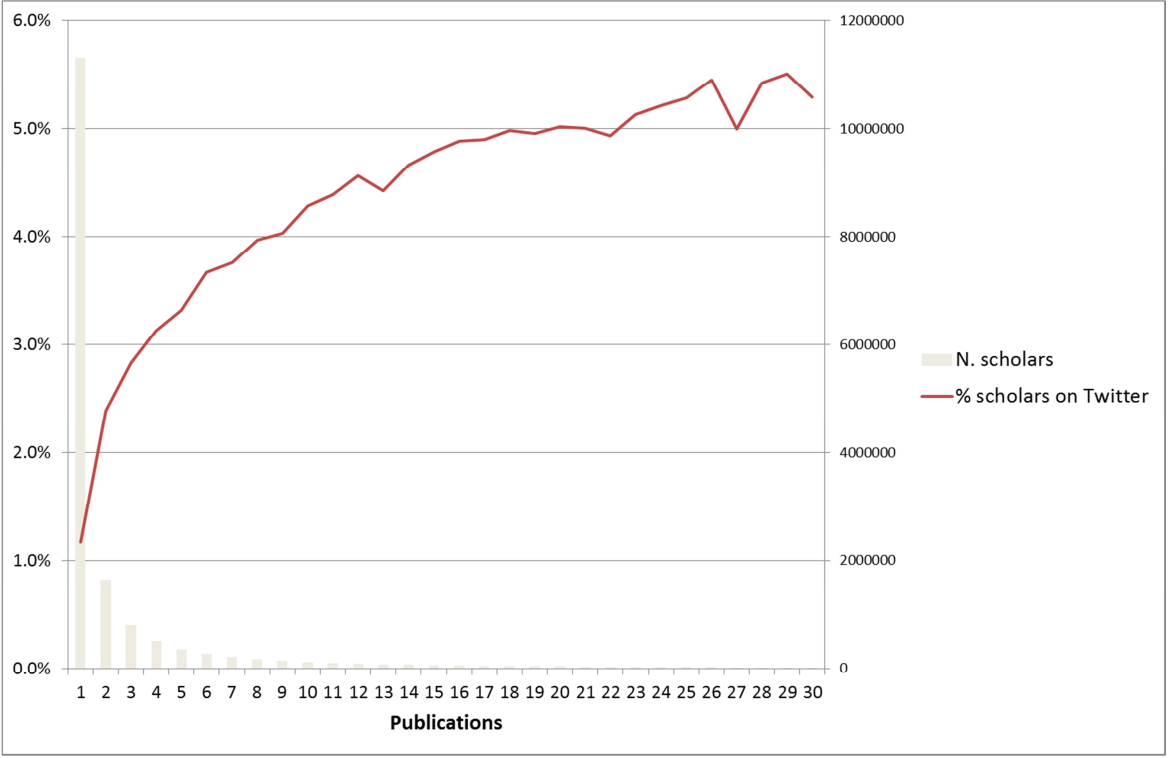

Figure 2. Percentage of scholars with a Twitter account by main domain of activity

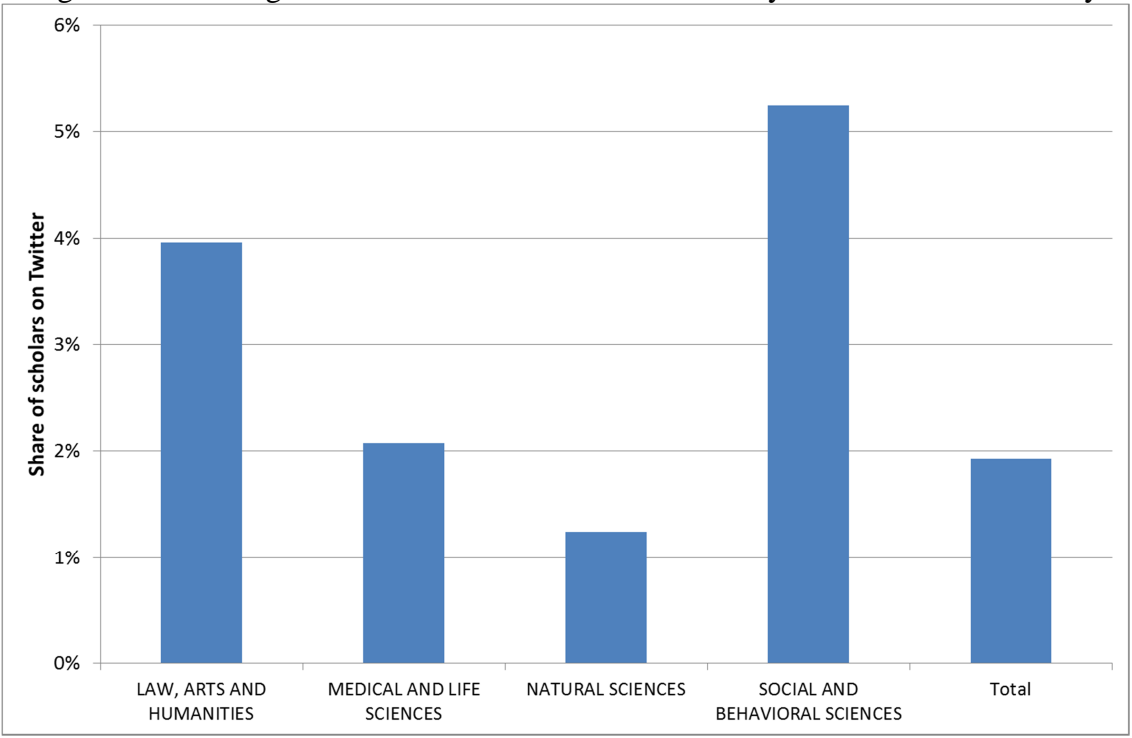

In Figure 3 an analysis on the age of the scholars is presented. In this case, age is determined by the year of first publication of the scholars. This year has been proven to be the best overall proxy of the academic age of scholars (Costas, Nane, & Larivière, 2015). Controlling for the number of publications we find that scientists with a Twitter account tend to be on average younger than those without a Twitter account. Not surprisingly, scholars with a lower number of publications tend to be younger overall than those with more publications, although the pattern of younger scholars on Twitter is observed for all groups of productivity.

Figure 3. Average Year of first publication (YFP) of scholars with a Twitter account vs. those without a Twitter account, controlling by the number of publications

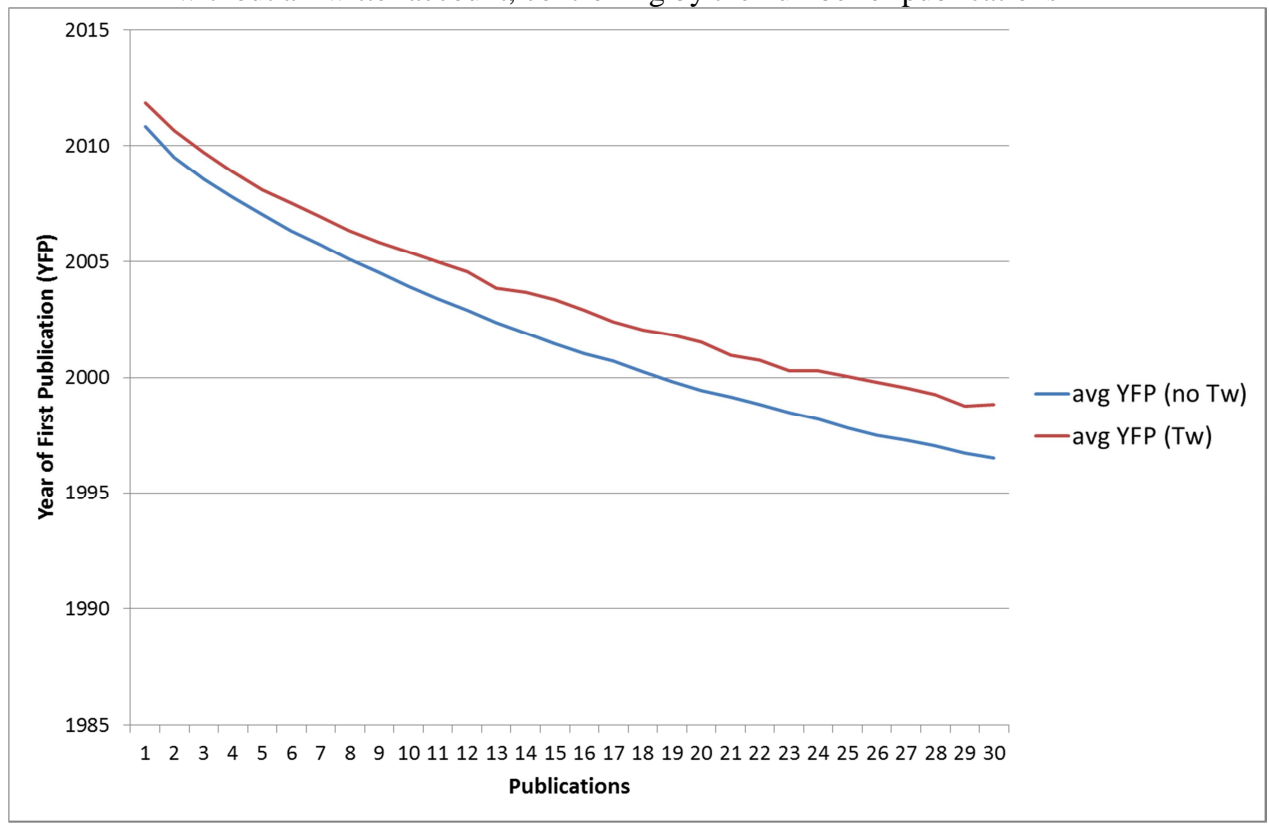

**Conclusions and further research**

In this paper we present a novel methodology for identifying scholars with a Twitter account. By combining bibliometric data from WoS and Twitter accounts from Altmetric.com we have obtained the largest matching of individual scholars with Twitter users made so far. Our methodology consists of a combination of matching algorithms, considering different linguistic elements of both author names and Twitter names; followed by a rule-based scoring system that weights the common occurrence of several elements related with the names, individual elements and activities of both Twitter accounts and scholars matched.

The methodology developed presents interesting advantages with respect to previous approaches:
- It is a systematic approach that can be applied to large sets of researchers. It relies on the role of scholars are publishing authors and is not dependent on aspects such as the biographical descriptions of the Twitter users or their presence in lists (cf. Ke, Ahn & Sugimoto, 2016).
- It is subject of validation by external gold standard, in this case a gold standard based on data provided by ORCID has been employed, but the analysis of other golden sets would be also possible.

- Once a Twitter account of an author is identified, it possible to link this with additional bibliometric data (e.g. affiliations, scientific domain, citation impact, collaboration patterns, etc.) related to the scientific author. It is of course also possible to extract data from the Twitter handle, thus being able to incorporate information on the online activity of the scholar (e.g. followers, followees, (re)tweeting activity, hashtags, etc.). This opens a unique possibility to exhaustive studies on the activities that scholars are performing in social media as well as in their publications.

The main limitation of this approach is its reliance on WoS (see also Ke, Ahn & Sugimoto, 2016) and Altmetric.com, which means it can only be used to identify scientists that publish in journals included in WoS. This means we would fail to identify a larger amount of humanities and social science scholars active on Twitter but not active in WoS. Similarly, scholars who are on Twitter but not tweeting scientific outputs (or tweeting outputs not properly tracked by Altmetric.com) would also be excluded. Also, activities in social media platforms different than Twitter are not considered in this analysis (e.g. the Chinese Weibo).
In any case, results show positive values in terms of precision and recall suggesting a strong validity of this methodology. More than 385,000 scholars can be linked to a Twitter account. In general, our results indicate a presence scholars on Twitter between 1% and 5% in the overall population of scholars in the WoS. The numbers of scholars on Twitter vary by levels of productivity, thus researchers with higher levels of production have also a stronger presence on Twitter, this very likely also caused by the reliance of the method on the number of publications of scholars. By disciplines we find a strong presence of researchers from the Social Sciences and the Humanities. Natural Sciences is the scientific domain with the lowest level of scholars on Twitter. Researchers on Twitter tend to be younger than those that are not on Twitter. These results align with those reported by Ke, Ahn & Sugimoto (2016) who also reported a higher presence of Social Sciences and Historians on Twitter, and lower levels of Life and Natural Sciences as well as Mathematicians. The high level of Humanist scholars is remarkable, considering that previous studies found that the Humanities is not one of the strongest scientific domains in altmetrics (Costas, Zahedi, & Wouters, 2015) as well as the fact that publications of scholars in this domain are least likely to be covered in the WoS.

The approach presented here also opens new and diverse venues of further research:
- Improvement of the matching and scoring algorithms. An important element that will need to be explore in the future is how to improve the matching methodology. For example, how to better match names in other non-Roman alphabets, the better parsing of Twitter names, abetter matching and scoring the pairs of authors and Twitter users, etc. All these are relevant research aspects that will need more attention in the future. Additionally, the creation and evaluation of additional gold standards (i.e. validated lists of authors and their Twitter and other social media accounts) will be a necessary step in order to validate future approaches of identification of scholars on social media.
- Incorporation of other databases. Including other bibiometric databases (e.g. Scopus, Google Scholar, Microsoft Academic) would increase the population of publications and scholars. Also, the inclusion of other altmetric data sources (e.g. PlumX Analytics) as well as working with social media platforms directly (as done in Ke, Ahn & Sugimoto, 2016) would increase the possibilities to identify scholars on social media platforms.
- Demography of scholars on Twitter. Based on the current research, further analysis should focus on more deeply studying the demography of scholars on Twitter. Thus, considering the combination of both bibliometric and Twitter data it will be possible to analyse the countries of the scholars on Twitter, their gender or their subdisciplines, among other perspectives (e.g. age, collaboration, mobility, etc.).

- Moreover, the combination of bibliometric and altmetric information opens a clear path to study of the relationship between bibliometric performance and Twitter and social media activity. Thus, questions related with the impact (citation and altmetric) of those scholars on Twitter compared to those who are not on Twitter will be key future questions.
- Similarly the analysis of the activities and interactions of scholars on Twitter will be an excellent method to better determine and contextualize the interactions that scholars are maintaining with other societal stakeholders, as suggested by Robinson-Garcia, van Leeuwen, & Rafols (2017).

It can be conclude that this approach represents as a qualitative first step towards a large identification of the science-Twitter universe in a systematic way combining both bibliometric and altmetric sources. A combination of our approach with the list approach (as in Ke, Ahn & Sugimoto, 2016) as well as an analysis of the biographies of the followers and followings of the identified scholars might help to counter the limitations of the current approach, opening the path to a more complete perspective of the true engagement of scholars on social media platforms.

**Acknowledgements**

This work has been supported by Eurostars-2 funded project SIA Graph. Rodrigo Costas was partially supported by funding from the DST-NRF Centre of Excellence in Scientometrics and Science, Technology and Innovation Policy (SciSTIP) (South Africa). The authors acknowledge the help by Josh Brown, Adèniké Deane-Pratt and Tom Deranville from ORCID in obtaining the gold standard database and the comments and feedback received from Cassidy Sugimoto and Vincent Larivière on early discussions of this paper.

**References**

Caron, E. & Van Eck, N. J. (2014). Large scale author name disambiguation using rule-based scoring and clustering. In E. Noyons (Ed.), *19th International Conference on Science and Technology Indicators. "Context counts: pathways to master big data and little data."* Leiden: CWTS-Leiden University.

Collins, K., Shiffman, D. & Rock, J. (2016). How are scientists using social media in the workplace? *PLoS ONE*, 11, 1–10.

Costas, R., Nane, T. & Larivière, V. (2015). Is the year of first publication a good proxy of scholars' academic age? In et al. Salah AA, Tonta Y, Akdağ Salah AA (Ed.), *Proceedings of the 15th international conference on scientometrics and informetrics* (pp. 988–998). Istanbul: Bogaziçi University Printhouse.

Costas, R., Zahedi, Z. & Wouters, P. (2015). The tematic orientation of publications mentioned on social media: large-scale disciplinary comparison of social media metrics with citatios. *Aslib Journal of Information Management*, 67, 260–288.

Haak, Laurel; Brown, Josh; Buys, Matthew; Cardoso, Ana Patricia; Demain, Paula; Demeranville, Tom; Duine, Maaike; Harley, Stephanie; Hershberger, Sarah; Krznarich, Liz; Meadows, Alice; Miyairi, Nobuko; Montenegro, Angel; Paglione, Laura; Pessoa, Lilian; Peters, Robert; Monge, Fran Ramírez; Simpson, Will; Wilmers, Catalina; Wright, Douglas (2016). *ORCID Public Data File 2016*. figshare. https://doi.org/10.6084/m9.figshare.4134027.v1

Hadgu, A. T. & Jäschke, R. (2014). Identifying and analyzing researchers on Twitter. *CEUR Workshop Proceedings*, 1226, 164–165.

Haustein, S., Bowman, T. D., & Costas, R. (2016). *Interpreting "altmetrics": viewing acts on social media through the lens of citation and social theories*. In C. R. Sugimoto (Ed.),

Theories of Informetrics: A Festschrift in Honor of Blaise Cronin (pp. 372–405). Berlin: De Gruyter Mouton. Retrieved from http://arxiv.org/abs/1502.05701

Haustein, S., Bowman, T. D., Holmberg, K., Peters, I., & Larivière, V. (2014). Astrophysicists on Twitter: An in-depth analysis of tweeting and scientific publication behavior. *Aslib Journal of Information Management*, 66(3), 279–296. http://doi.org/10.1108/AJIM-09-2013-0081

Holmberg, K. & Thelwall, M. (2014). Disciplinary differences in Twitter scholarly communication. *Scientometrics*, 101, 1027–1042.

Hwong, Y.-L., Oliver, C., Van Kranendonk, M., Sammut, C. & Seroussi, Y. (2016). What makes you tick? The psychology of social media engagement in space science communication. *Computers in Human Behavior*, 68, 480–492.

Ke, Q., Ahn, Y.-Y. & Sugimoto, C. R. (2016). A Systematic Identification and Analysis of Scientists on Twitter. *PLoS ONE*, 1–28. Retrieved from http://arxiv.org/abs/1608.06229

Larivière, V. & Costas, R. (2016). How Many Is Too Many? On the Relationship between Research Productivity and Impact. *PloS One*, 11, e0162709.

Lulic, I. & Kovic, I. (2013). Analysis of emergency physicians' Twitter accounts. *Emergency Medicine Journal : EMJ*, 30, 371–6.

Mislove, A., Lehmann, S., Ahn, Y., Onnela, J. & Rosenquist, J. N. (2011). Understanding the Demographics of Twitter Users. *Artificial Intelligence*, 554–557. Retrieved from http://www.aaai.org/ocs/index.php/ICWSM/ICWSM11/paper/viewFile/2816/3234

Mondragon, N. I., Gil de Montes, L. & Valencia, J. (2017). Ebola in the Public Sphere. *Science Communication*, 39, 101–124.

Paglione, Laura; Peters, Robert; Wilmers, Catalina; Simpson, Will; Montenegro, Angel; Ramírez Monge, Fran; Tyagi, Shobhit; Krznarich, Elizabeth; Demeranville, Tom; Brown, Josh; Miyairi, Nobuko; Buys, Matthew; Cardoso, Ana; Sethate, Cheryl; Haak, Laurel (2015). *ORCID Public Data File 2015*. figshare. https://doi.org/10.6084/m9.figshare.1582705.v1

Ross, C., Terras, C., Warwick, M. & Welsh, A. (2011). Enabled Backchannel: Conference Twitter use by Digital Humanists. *Journal of Documentation*, 67, 214–237.

Rowlands, I., Nicholas, D., Russell, B., Canty, N. & Watkinson, A. (2011). Social media use in the research workflow. *Learned Publishing*, *24*(3), 183–195. http://doi.org/10.1087/20110306

Sharma, N. K., Ghosh, S., Benevenuto, F., Ganguly, N. & Gummadi, K. (2012). Inferring who-is-who in the Twitter social network. *ACM SIGCOMM Computer Communication Review*, *42*, 533.

Sloan, L., Morgan, J., Burnap, P. & Williams, M. (2015). Who tweets? deriving the demographic characteristics of age, occupation and social class from Twitter user meta-data. *PLoS ONE*, 10, 1–20.

Sloan, L., Morgan, J., Housley, W., Williams, M., Edwards, A., Burnap, P. & Rana, O. (2013). Knowing the Tweeters. *Sociological Research Online*.

Sugimoto, C. Work, S., Larivière, V. & Haustein. (2016). Scholarly use of social media and altmetrics: review of the literature. Retrieved from https://arxiv.org/abs/1608.08112.

Van Noorden, R. (2014). Online collaboration: Scientists and the social network. *Nature*, 512, 126–129.

Veletsianos, G. (2012). Higher education scholars' participation and practices on Twitter. *Journal of Computer Assisted Learning*, 28, 336–349.

Veletsianos, G. & Kimmons, R. (2016). Scholars in an increasingly open and digital world: How do education professors and students use Twitter? *Internet and Higher Education*, 30, 1–10.